# On the Capacity of Multi-Hop Wireless Networks with Partial Network Knowledge

Alireza Vahid
Cornell University
Ithaca, NY, USA.
av292@cornell.edu

Vaneet Aggarwal
Princeton University
Princeton, NJ, USA.
vaggarwa@princeton.edu

A. Salman Avestimehr
Cornell University
Ithaca, NY, USA.
avestimehr@ece.cornell.edu

Ashutosh Sabharwal
Rice University
Houston, TX, USA.
ashu@rice.edu

*Abstract*—In large wireless networks, acquiring full network state information is typically infeasible. Hence, nodes need to flow the information and manage the interference based on partial information about the network. In this paper, we consider multi-hop wireless networks and assume that each source only knows the channel gains that are on the routes from itself to other destinations in the network. We develop several distributed strategies to manage the interference among the users and prove their optimality in maximizing the achievable normalized sum-rate for some classes of networks.

## I. Introduction

One of the key challenges in the design of wireless networks is optimizing system efficiency while the network is constantly in flux. To manage system resources efficiently, it is crucial to keep track of the state of the network and determine what resources are actually available. However, in large wireless networks, keeping track of the state for making optimal decisions is typically infeasible. Thus, in the absence of centralization of network state information, nodes have limited local view of the network and make distributed decisions, which are based on their own local view of the network. The key question then, is how do optimal distributed decisions perform in comparison to the optimal decisions when full network state information is available at all nodes.

All wireless networks (eg. 2G/3G, WiFi, WiMax Bluetooth, ZigBee) rely on a host of distributed coordination protocols for their operation. The numerous protocol innovations have led to many practical advances in deployed networks. However, there is very little work in understanding the fundamental limits for distributed protocols [1, 2].

A systematic study to understand the role of limited network knowledge, was first initiated in [3, 4] for single-layer networks, where the authors used a message-passing abstraction of network protocols to formalize the notion of limited network view at each node in the form of number of message rounds; each message round adds two extra hops of channel information at each source. The key result was that distributed decisions can be either sum-rate optimal or can be arbitrarily worse than the global-information sum-capacity.

This result was further strengthened for arbitrary $k$-user single-layer interference network in [5, 6, 8], where the authors proposed a new metric, normalized sum-capacity, to measure the performance of distributed decisions. Further, the authors computed the capacity of distributed decisions for several network topologies with one-hop and two-hop network information at each source. In this paper, we characterize the next major step in understanding the performance of distributed decisions for general acyclic linear deterministic networks [7].

In single-layer networks, the main challenge is to manage the interference among the sources with partial network information. However, in more general networks, we face an additional complication - the relaying of information with partial network information in addition to managing interference at all nodes. While the hop-count model of [6] can still be used to quantify partial network-state information, we will develop a more scalable model for general acyclic networks in this paper, which is based on the knowledge about the routes in the network. The motivation for this model is that in general networks, coordination protocols like routing, aim to discover source-destination routes in the network. Hence, a more suitable quanta for network-state information is the number of such end-to-end routes that are known at the nodes.

Our results are three-fold. First, as mentioned we will develop a more scalable model for general acyclic networks in this paper. Second, we develop new strategies that can be used with this local knowledge, and third, we prove the optimality of these strategies for a variety of networks.

In this paper, we will consider the case where each source has full information about network topology, as well as full information about the routes from itself to all the destinations. Each node (which is not a source), has the union of the information of all those sources that have a route to it. This model of partial information will be formally defined in Section II-B. We will start by a natural extension of Maximum Independent Set (MIS) scheduling [5, 6], called the Maximum Independent Route (MIR) scheduling and we notice that it fails to perform optimally in many cases. We will remove a main limitation of MIR scheduling, and through this, we develop the Maximum Independent Link (MIL) scheduling and we prove its optimality for a class of networks, namely $k \times 2 \times k$ networks. We will face an interesting example, that shows the requirement of network coding at relays. This motivates us to consider a broader class of networks and strategies.

The rest of the paper is organized as follows. In section II, we will introduce our network model and the new model to capture partial network knowledge. Section III is dedicated to a brief overview of the results for single-layer networks. In Section IV, we describe our main results including the achievability schemes. We introduce upper bounds on the performance of distributed decisions based on local network knowledge in Section V. Finally, Section VI concludes the paper and presents some future directions.

## II. PROBLEM FORMULATION

In this section, we first introduce our channel and network models. We will then describe our model for partial network knowledge. Finally, we define *normalized sum-rate* and *normalized sum-capacity* with partial network knowledge.

### A. Network Model

We represent a network with a directed graph $\mathsf{G} = (\mathsf{V}, \mathsf{E}, \{\mathsf{n}_{ij}\}_{(i,j)\in\mathsf{E}})$, where $\mathsf{V}$ is the set of vertices representing nodes in the network and $\mathsf{E}$ is the set of edges representing links among the nodes, and $\{\mathsf{n}_{ij}\}_{(i,j)\in\mathsf{E}}$ is the set of channel gains that will be described later in this section. We further assume that $\mathsf{G}$ is acyclic, and there is no outgoing edge from any destination and no incoming edge to any source. We denote the vertices in $\mathsf{G}$ as $\mathsf{N}_i$'s, $i = 1, 2, \ldots, |\mathsf{V}|$. Of the nodes in the network, there are $k$ source-destination pairs and we label them as $(\mathsf{S}_i, \mathsf{D}_i)$, $i = 1, 2, \ldots, k$. The rest of the nodes are considered as relays who try to facilitate the communication between sources and destinations.

A *route* from $\mathsf{S}_i$ to $\mathsf{D}_j$ is simply a path from $\mathsf{S}_i$ to $\mathsf{D}_j$ in $\mathsf{G}$. The *in-degree* of a node $\mathsf{N}_i$, denoted by $\mathsf{d}_i^{in}$, is the number of in-coming edges connected to it. Similarly, the *out-degree* of a node $\mathsf{N}_i$, denoted by $\mathsf{d}_i^{out}$, is the number of out-going edges connected to it, see Figure 1. We will explicitly distinguish between the in-degree and the out-degree of a node, and will not use sum of the two throughout the paper.

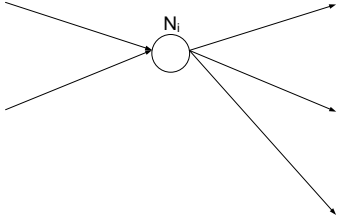

Fig. 1. A node with in-degree 2 and out-degree 3.

We consider the linear deterministic model [7] for the channels in the network. In this model, there is a non-negative integer, $n_{ij}$, associated with each link $(i, j) \in \mathsf{E}$, which represents its gain. Let $q$ be the maximum of all the channel gains in this network. In the linear deterministic model, the channel input at node $i$ at time $t$ is denoted by $X_i[t] = [X_{i_1}[t], X_{i_2}[t], \ldots, X_{i_q}[t]]^T \in \mathbb{F}_2^q$. The received signal at node $j$ at time $t$ is denoted by $Y_j[t] = [Y_{j_1}[t], Y_{j_2}[t], \ldots, Y_{j_q}[t]]^T \in \mathbb{F}_2^q$, and is given by

$$Y_j[t] = \sum_{i:(i,j)\in\mathsf{E}} \mathbf{S}^{q-n_{ij}} X_i[t], \quad (1)$$

where $\mathbf{S}$ is the $q \times q$ shift matrix and the operations are in $\mathbb{F}_2^q$. If a link between $\mathsf{N}_i$ and $\mathsf{N}_j$ does not exist, we set $n_{ij}$ to be zero.

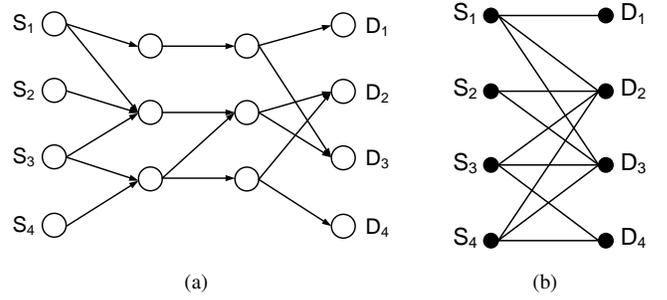

Fig. 2. *(a) A multi-layer acyclic network, and (b) its route-adjacency graph.*

### B. Modeling of Partial Network Knowledge

In this subsection, we will describe a model for partial network information that will be used in this paper.

We first define the *route-adjacency graph* $\mathsf{P}$, which is an undirected bipartite graph, consisting of all the sources on one side and all the destinations on the other side, see Figure 2 for a depiction. A source $\mathsf{S}_i$ and a destination $\mathsf{D}_j$ are adjacent in $\mathsf{P}$, if there exists a route between them in $\mathsf{G}$. Each link $(\mathsf{S}_i, \mathsf{D}_j)$ in $\mathsf{P}$ is assigned with the set of all channel gains that are in a route from $\mathsf{S}_i$ to $\mathsf{D}_j$.

We now define the partial network knowledge that will be used in the paper as the following:

- All nodes have full knowledge of the network topology, $(\mathsf{V}, \mathsf{E})$, (i.e., which links are in $\mathsf{G}$, but not their channel gains) (this side information is denoted by side information SI),
- Each source, $\mathsf{S}_i$, has full knowledge of the channel gains assigned to those edges in $\mathsf{P}$ that are at most $h$ hops away from $\mathsf{S}_i$ (this local knowledge at a source is denoted by $L_{\mathsf{S}_i}$),
- Each node $\mathsf{N}_i$ (which is not a source) has the union of the information of all those sources that have a route to it (this local knowledge at node is denoted by $L_{\mathsf{N}_i}$).

We name this model for partial network knowledge as $h$-*local view*. Note that this model is a generalization of the hop-based model for partial network knowledge in single layer networks [6]. In this paper, we focus on the case where $h = 1$.

### C. Normalized Sum-Capacity

We now describe our metric for evaluating network capacity with partial network knowledge. As in [6, 8], we use the *normalized sum-capacity*, which represents the fraction of the sum-capacity with full knowledge that can be achieved when nodes have only partial knowledge about the network.

For each source $\mathsf{S}_i$, let message $\mathsf{W}_i$ be encoded as $X_{\mathsf{S}_i}^n$, where $n$ is the block length, using the encoding function $e_i(\mathsf{W}_i | L_{\mathsf{S}_i}, \mathsf{SI})$, which depends on the available network knowledge, $L_{\mathsf{S}_i}$, and the side information, SI. Each relay in the network creates its input to the channel $X_{\mathsf{N}_i}$, using the encoding function $f_{\mathsf{N}_i}(Y_{\mathsf{N}_i} | L_{\mathsf{N}_i}, \mathsf{SI})$, which depends on the available network knowledge, $L_{\mathsf{N}_i}$, and the side information, SI. A relay strategy is defined as the union of of all encoding functions in the network, $\{f_{\mathsf{N}_i}(Y_{\mathsf{N}_i}|L_{\mathsf{N}_i},\mathsf{SI})\}$. Destination $\mathsf{D}_i$ is only interested in decoding $\mathsf{W}_i$ and it will decode the message using the decoding function $\hat{\mathsf{W}}_i = d_i(Y_{\mathsf{D}_i}^n|L_{\mathsf{D}_i},\mathsf{SI})$, where $L_{\mathsf{D}_i}$ is the destination's network knowledge. A strategy is defined as the union of of all encoding and decoding functions in the network, $\{e_i(\mathsf{W}_i|L_{\mathsf{S}_i},\mathsf{SI}), d_i(Y_{\mathsf{D}_i}^n|L_{\mathsf{D}_i},\mathsf{SI})\}, i=1,2,\ldots,k$, and the relay strategy. We note that the local view can be different from node to node.

An error occurs when $\hat{\mathsf{W}}_i \neq \mathsf{W}_i$ and we define the decoding error probability, $\lambda_i$, to be equal to $P(\hat{\mathsf{W}}_i \neq \mathsf{W}_i)$. A rate tuple $(R_1, R_2, \ldots, R_k)$ is said to be achievable, if there exists a strategy such that the decoding error probabilities $\lambda_1, \lambda_2, \ldots, \lambda_k$ go to zero as $n$ goes to infinity for all network states consistent with the side information. The sum-capacity $C_{sum}$, is the supremum of $\sum_i R_i$ over all possible encoding and decoding functions with full network knowledge. We will now define the normalized sum-rate and the normalized sum-capacity.

**Definition 1** ([6]). **Normalized sum-rate** *of $\alpha$ is said to be achievable for a set of network states with partial information if there exists a strategy such that following holds. The strategy yields a sequence of codes having rates $R_i$ at the source $i$ such that the error probabilities at the destinations, $\lambda_1(n), \cdots \lambda_K(n)$, go to zero as $n$ goes to infinity, satisfying*

$$\sum_i R_i \geq \alpha C_{sum} - \tau$$

*for all the network states consistent with the side information, and for a constant $\tau$ that is independent of the channel gains, but may depend on the side information $\mathsf{SI}$. Here $C_{sum}$ is the sum-capacity of the whole network with the full information.*

**Definition 2** ([6]). **Normalized sum-capacity** $\alpha^*$, *is defined as the supremum of all achievable normalized sum-rates $\alpha$. Note that $\alpha^* \in [0,1]$.*

## III. BACKGROUND ON SINGLE-LAYER NETWORKS

In this section, we will summarize the results in [6, 8], where the increase of normalized sum-capacity with increasing local view was considered. The interference network has $h$-local view if each source knows the channel gains of all the links which are at-most $h$ hops distant from it, and each destination knows the channel gains of all the links that are at-most $h+1$ hops distant from it. The local view defined in this paper reduces to that in [6] for single-layer networks.

With $h$-local view, one intuitive solution is for nodes to coordinate their transmissions such that the nodes beyond $h$ hops transmit only if they can cause no interference with $h$-hop size sub-network and thus each connected sub-network operates as if it is a network with full global information. This is formalized through the notion of an independent graph, which is defined as a sub-graph which admits a distributed encoding and decoding scheme which achieves same sum-capacity as a scheme with full global information. This intuition was used in [6] to propose *maximal independent graph (MIG)* scheduling, where the network is divided into sub-graphs (equivalently sub-networks) and the sub-graphs are scheduled orthogonally over time. The sub-graphs are chosen such that they are maximal independent graphs which ensures highest spatial reuse of the users. The authors also showed that MIG scheduling is optimal in many cases, but is not always optimal. The rest of this section will describe the results in [6] with 1-local view for a linear deterministic model. In this section, we will define MIS scheduling, which is a special case of MIG scheduling with 1-local view. We will further describe an example where MIS scheduling is not optimal, and define CS scheduling algorithm as an achievable strategy. Finally, we will describe the optimal strategy that can be used with 1-local view for which the optimal strategy for a special case of interference network is required as input.

### A. MIS Scheduling

For 1-local view, the MIG scheduling strategy reduces to maximal independent set (MIS) scheduling that can be described as follows. An independent set $A_i \subseteq \{1,\cdots,k\}$ is a set that contains mutually non-interfering nodes. A maximal independent set (MIS) is an independent set $A_i$ such that $A_i \cup \{x\}$ is not an independent set for any $x \in \{1,\cdots,k\}\setminus A_i$. Using $t$ time-slots, a maximal independent set $A_i$ is scheduled in each time-slot such that

$$\min_i \frac{1}{t}\sum_{j=1}^t \mathbf{1}_{i\in A_j}$$

is maximized over the choice of $t$ and $A_1\cdots,A_t$. When a user is scheduled, it sends at the direct channel rate. The resulting strategy achieves a normalized sub-rate of $\alpha = \min_i \frac{1}{t}\sum_{j=1}^t \mathbf{1}_{i\in A_j}$.

This is similar to the following vertex coloring algorithm. To relate to vertex coloring, we will need the concept of conflict graph [9] derived from $\mathsf{G}$ as follows. Consider a graph $\mathsf{C}$ with $k$ vertices (half as many as present in $\mathsf{G}$), where two vertices $i$ and $j$ are connected if there is an edge between $\mathsf{S}_i$ and $\mathsf{D}_j$ or between $\mathsf{S}_j$ and $\mathsf{D}_i$ in $\mathsf{G}$. Suppose that there are $t$ colors, labeled $1, 2, \cdots, t$. We assign $m \leq t$ of these colors to each vertex in $\mathsf{C}$ such that the sets of colors associated with two vertices connected by an edge are disjoint. In conventional graph coloring [10], each vertex has only one color and the objective is to assign a color to each vertex such that adjoining vertices have different colors. In contrast, the generalized *set coloring* algorithm can assign multiple colors to each vertex as long as the color sets for adjoining vertices are disjoint. The best set coloring corresponds to MIS schedule and maximizes $\frac{m}{t}$ with $m$ and $t$ as variables. The scheduling algorithm uses $t$ time slots and schedules the vertices with color $i$ in the $i^{th}$ time-slot.

An $m$-fold coloring of a graph is an assignment of sets of size $m$ to vertices of a graph such that adjacent vertices receive disjoint sets. A $t : m$-coloring is a $m$-fold coloring out of $t$ available colors. The $m$-fold chromatic number $\xi_m$ is the least $t$ such that a $t : m$-coloring exists. Note that MIS Scheduling achieves $\alpha = \max_{m\in\mathbb{N}} \frac{m}{\xi_m}$, where $\xi_m$ is the $m$-fold chromatic number of the conflict graph. The following theorem gives an optimality condition of MIS Scheduling algorithm in terms of the $m$-fold chromatic number of the conflict graph.

**Theorem 1** ([6])**.** *If the conflict graph of a single-layer interference network has $m$-fold chromatic number of at most $2m$ for some $m \in \mathbb{N}$, then the MIS scheduling algorithm is optimal, i.e. achieves normalized sum-capacity with $1$-local view.*

In [6], MIS scheduling was shown to be optimal in several cases as described next.

**Theorem 2** ([6])**.** *MIG scheduling is optimal with $1$-local view for the following single-layer networks.*

1) *All the three-user interference networks, except the 3-user folded-chain,*
2) *chain interference network,*
3) *d-to-many interference network,*
4) *many-to-d interference network,*
5) *fully-connected interference network.*

*For all cases, the achievability holds for $\tau = 0$.*

### B. Coded Set Scheduling

In this subsection, we will describe an example where MIS scheduling is not optimal and further define a new achievable strategy called coded set (CS) scheduling.

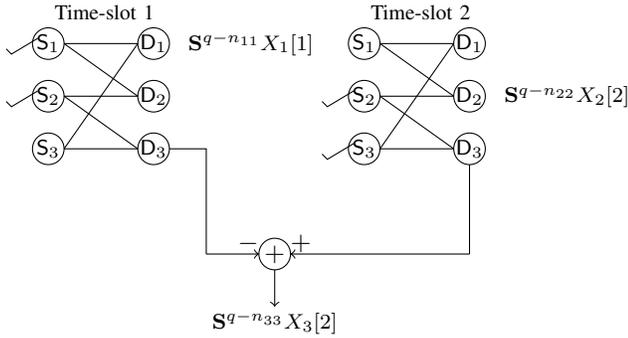

Fig. 3. Two time-slots for CS scheduling. The sources with a tick sign transmit, the second user repeats $X_2$ ($X_2[1] = X_2[2]$) in the two time-slots.

We will now illustrate the only case when MIS Scheduling is not optimal in a 3-user interference network, which is a folded-chain interference network (this network was called "cyclic Z-chain" in [6], however in order to avoid any confusion about our intuition of cyclic networks, we use the name "folded-chain"). The MIS scheduling algorithm uses three time-slots, scheduling user $i$ in time-slot $i$. Thus, MIS scheduling achieves $\alpha = \frac{1}{3}$ (note that there are only 3 independent sets consisting of individual users and thus optimality of $\frac{1}{3}$ using MIS scheduling is straightforward).

We will now describe another strategy for this example, which uses two time slots as follows (and depicted in Figure 3). The main idea is to perform coding across time. In the first time slot, we schedule $A_1 = \{1, 2\}$ and in second time slot, we schedule $A_2 = \{2, 3\}$ such that the codeword of the second user is repeated in the two time slots. All the users send at the rate equal to the direct link capacity to the intended destination ($n_{ii}$). We will now show that the data can be decoded at the intended destinations. The first destination can decode its data in the first time slot since it receives no interference. The second destination can similarly decode the data in the second time slot. The third destination on the other hand subtracts the data received in the first time slot from that in the second time slot, which gives an interference-free direct signal that can be decoded. Thus, all the destinations can decode the data and this strategy achieves $\alpha = 1/2$. This result, will be formally generalized as follows.

**Definition 3.** *A $(k, d)$ **folded-chain network**, $d \leq k$, is an interference network with $k$ source-destination pairs. Source $i$, is connected to destinations $(i, i+1, \ldots, i+d)$ **mod** $k$, $i = 1, 2, \ldots, k$. Note that in this networks, all in-degrees and all out-degrees are equal to $d$.*

**Theorem 3.** *The normalized sum-capacity of a $(k, d)$ folded-chain network with $1$-local view is*

$$\alpha^* = \begin{cases} 1 & k = 1, 2 \\ \frac{1}{d} & k \geq 3 \end{cases} \quad (2)$$

*Proof:* The result for $k = 1, 2$ is trivial. For $k \geq 3$, we first establish an upper bound on the achievable normalized sum-rate. Consider a $(k, d)$ folded-chain network where the channel gain of a link from source $i$ to destinations $i, i+1, \ldots, d$ is equal to $n$, $i = 1, 2, \ldots, d$, and all the other channel gains are equal to zero. See Figure 4 for a depiction.

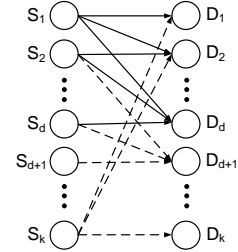

Fig. 4. *Channel gain assignment in a $(k, d)$ folded-chain network. All solid links have capacity $n$ and all dashed links have capacity $0$.*

Suppose, a normalized sum-rate of $\alpha$ is feasible in this network with 1-local view. Each source due to its local view of the network, should transmit at a rate $\geq \alpha n - \tau$, to guarantee a normalized sum-rate of $\alpha$. Destination 1 receives its message, $W_1$, with no interference and can decode it. Destination 2, decodes its message and removes it from the received signal, what is left is exactly the same as what destination 1 receives, therefore destination 2 is able to decode $W_1$ and $W_2$. If we continue this argument, we see that destination $d$ should be able to decode all $W_i$'s, $i = 1, 2, \ldots, d$. From the MAC capacity at destination $d$ we have

$$d(\alpha n - \tau) \leq n \Rightarrow (d\alpha - 1)n \leq d\tau \quad (3)$$

since this has to hold for all values of $n$ where $\alpha$ and $\tau$ are independent of $n$, we get $\alpha \leq \frac{1}{d}$.

The achievability scheme works as follows. First, assume that $d < k < 2d$, and we will later generalize the achievability scheme for arbitrary $k$. Let $d' = k - d + 1$. In time slot $j$, sources $j, (j+1), \ldots, (j + d' - 1)$ transmit their messages at full rate $n$, for $j = 1, 2, \ldots, d$. This way, in time slot $j$, destination $j$ will get its message interference-free and as a result, first $d$ destinations can decode their messages easily.

Consider destination $k$. In time slot 1, it will receive interference from $\mathsf{S}_{d'}$. During second time slot, it will receive interference from $\mathsf{S}_{d'}$ and $\mathsf{S}_{d'+1}$. It will remove the signal received during first time slot, from the new signal, and it has access to the interference coming from $\mathsf{S}_{d'+1}$. Similarly, it will have access to the interference it receives from any source. Finally at time slot $d$, it will remove all the interference signals to get its message, $\mathsf{W}_k$, interference-free.

A similar argument is valid for any destination. They will successively remove the signal that they received in previous time slot from their new signal to have access to the interference. Through this scheme, destination $i$, $i = d+1, d+2, \ldots, k$, will be able to remove interference from its message and decode it.

For general $k$ of the form $c(2d-1) + m$, where $c \geq 2$ and $0 \leq m < (2d-1)$, we implement the scheme for source-destination pairs $1, 2, \ldots, 2d-1$ as if they are the only pairs in the network. The same for source-destination pairs $2d, 2d+1, \ldots, 4d-2$ and etc. Finally, for the last $m$ source-destination pairs, we implement the scheme with $d' = \max\{m-d+1, 1\}$. Note that in the case where $m \leq d$, we have $d' = 1$, and we turn on source $c(2d-1) + j$ in time slot $i$, where $j = ((i-1) \bmod m) + 1$. This proves that $\alpha^*$ as in the statement of the theorem is achievable. Also note that this achievability scheme has $\tau = 0$. ∎

**Remark:** If all nodes follow the same strategy, *i.e.* successively remove received signal from the new one, they will have access to all interference signals.

The achievability scheme developed for the $(k, d)$ folded-chain network with 1-local view, motivates an extension of the MIS Scheduling algorithm to involve coding. This new scheduling algorithm is called Coded Set Scheduling (CS Scheduling) [6]. For this, the authors consider subgraphs $A \subseteq G$ with a set of sources $\mathsf{S}_i$ and all the destinations $\{\mathsf{D}_1, \cdots, \mathsf{D}_k\}$ in the subgraph so as to not throw away any received signal. Suppose that each source generates $k$ independent codewords (The rate of these codewords will be $n_{ii}$). Let $M_{i,j}$ be a vector of time-slots in which source $\mathsf{S}_i$ is transmitting the $j^{th}$ codeword. Note that each time-slot should be used at a source $\mathsf{S}_i$ for only one codeword, thus giving $M_{i,u}$ and $M_{i,v}$ disjoint for $u \neq v$. Thus, in time-slot $u$, the subgraph $A_u$ used has sources $\mathsf{S}_i$ where $i$ satisfies $M_{i,j} \supseteq \{u\}$ for some $1 \leq j \leq k$. The sets $M_{ij}$ and thus $A_u$, $t$ and $k$ are all design variables for the CS scheduling algorithm that satisfy some conditions on the constraint matrix, which is defined next.

We form a binary constraint matrix $F_i$ of size $kd_i \times t$ at each destination $i$ which is defined as follows. The constraint matrix has $d_i$ blocks of size $k \times t$ where the top block corresponds to the transmitted signal from $\mathsf{S}_i$ while the rest belong to the different sources causing interference at $\mathsf{D}_i$. In each $k \times t$ subpart of this matrix, only the entries $(j, M_{i,j})$ are 1 for all $1 \leq j \leq k$. Suppose that the $t$ columns of the constraint matrix are denoted as $\mathsf{Q}_{i,1}, \cdots \mathsf{Q}_{i,t}$ respectively. Suppose that a $kd_i \times t$ matrix with the top $k \times k$ part as an identity and rest of the elements 0 can be formed by choosing each column $j$ as $\sum_{l=1}^{t} a_{jl} \mathsf{Q}_{i,l}$ where $a_l$'s are binary and addition is binary addition. If such a transformation exist at destination $i$, this configuration is feasible at vertex $i$. If the assignment of $M_{ij}$ is feasible at each vertex, this strategy achieves $\alpha$ of $k/t$. The strategy that achieves the maximum $k/t$ is called Coded Set (CS) Scheduling.

The scheduling algorithm uses $t$ time-slots. Each user forms $k$ independent codewords at rate $n_{ii}$. User $i$ transmits codeword $j$ in time-slots corresponding to $M_{i,j}$. It is easy to see that the data can be decoded at the destinations. The constraint matrix reduction represents that all the $k$ independent codewords can be decoded in the presence of the interference from other sources.

### C. Optimal Approach

We note that MIS is not always optimal. In this subsection, we describe the optimal algorithm for single-layer networks with 1-local view.

**Definition 4** ([6]). *A **binary model** of a given interference network is a linear deterministic model with channel gains of links in* $\mathsf{E}$ *equal to* 1 *and all the rest equal to* 0.

**Definition 5** ([6]). **Symmetric capacity** *of an interference network is the maximum $r$ such that rate pair $(r, r, \cdots, r)$ is in the capacity region of the interference network with the full information of network and channel gains.*

The authors of [6] showed that the normalized sum-capacity of a given interference network with 1-local view is the symmetric capacity of the binary model of that interference network as described below.

**Theorem 4** ([6]). *The normalized sum-capacity of a given interference network with 1-local view is the symmetric capacity of the binary model of that interference network.*

The achievable strategy is to use the symmetric capacity achieving scheme for the binary model of interference network at all the bit levels of the original linear deterministic interference network.

## IV. TRANSMISSION STRATEGIES FOR MULTI-LAYER NETWORKS WITH PARTIAL INFORMATION

In this section, we focus on 1-local view at the nodes as defined in Section II-B. We will introduce a set of successively more efficient transmission strategies that only require 1-local view and achieve the normalized sum-capacity for several multi-layer networks.

### A. Maximum Independent Route (MIR) Scheduling

We start this section by considering a natural extension of the MIS scheduling defined in section III. We apply

MIS scheduling to the route-adjacency graph of a general network, and we label this achievability strategy as Maximum Independent Route (MIR) scheduling.

We now go through an example to illustrate MIR scheduling. Consider the network depicted in Figure 5(a) with 1-local view. The route-adjacency graph of this network is also depicted in Figure 5(b). Applying MIS scheduling to the graph in Figure 5(b), we can schedule pairs 1 and 3 to communicate at full rate in one time slot and pair 2 at a different time slot, which results in a normalized sum-rate of $\alpha = \frac{1}{2}$. This strategy in the original network of Figure 5(a), means to turn on source-destination pairs 1 and 3 and all the links that are on a route for either of them in one time slot, and the same for source-destination pair 2 in a different time slot. We show that this is indeed the normalized sum-capacity of this network with 1-local view, and as a result MIR scheduling is optimal in this case.

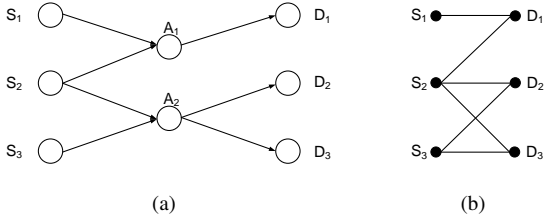

Fig. 5. (a) A network where MIR scheduling is optimal, and (b) its route-adjacency graph.

**Theorem 5.** *The normalized sum-capacity of the network in Figure 5(a) with* 1*-local view, is* $\alpha^* = \frac{1}{2}$ *and is achieved by MIR scheduling.*

*Proof:* We have already shown that $\alpha = \frac{1}{2}$, is achievable through MIR scheduling. Now we prove the converse by assuming that all the links in the network have equal capacity of $n$ and assume that a normalized sum-rate of $\alpha$ is achievable. Since each source has 1-local view and it has a capacity of $n$ to its destination when other source-destination pairs are silent, it has to transmit at a rate $\geq \alpha n - \tau$, to guarantee a normalized sum-rate of $\alpha$. This is due to the fact that from its point of view, it is possible that other source-destination pairs have capacity zero. Relay $A_2$, has all the information that destinations 2 and 3 have, and it should be able to decode $W_2$ and $W_3$ using its $n$ signal levels. The MAC capacity at relay $A_2$ results in

$$2\alpha n - 2\tau \leq n \Rightarrow (2\alpha - 1)n \leq 2\tau \quad (4)$$

since this has to hold for all values of $n$ where $\alpha$ and $\tau$ are independent of $n$, we get $\alpha \leq \frac{1}{2}$. Note that the achievability scheme holds for $\tau = 0$. ∎

While MIR scheduling performed optimally in the example of Figure 5, we now illustrate one of its main deficiencies. Consider the network depicted in Figure 6(a) with 1-local view. Applying MIS scheduling to its route-adjacency graph as depicted in Figure 6(b), we achieve a normalized sum-rate of $\alpha = \frac{1}{3}$. However, we show that the normalized sum-capacity of this network with 1-local view, is $\alpha^* = \frac{1}{2}$, and as a result MIR scheduling is *not* optimal in this case.

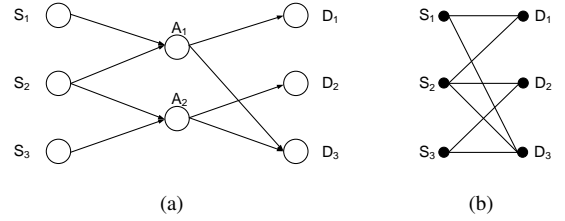

Fig. 6. (a) A network where MIR scheduling is not optimal, and (b) its route-adjacency graph.

The argument for the converse of Theorem 5 is valid here, *i.e.* $\alpha \leq \frac{1}{2}$. To achieve this normalized sum-rate, we will schedule nodes separately in each layer. In the first layer, we schedule Sources 1 and 3 to transmit at full rate in the first time slot and Source 2 in the second time slot. In the second layer, we schedule relays to serve destinations 1 and 2 at the same time and Destination 3 in a different time slot. This way, each source-destination pair achieves its maximum possible rate in two time slots, and we have $\alpha = \frac{1}{2}$.

The previous example illustrated a major deficiency of MIR scheduling, which is the fact that we apply the same scheduling to all nodes in a route and we ignore the flexibility of scheduling nodes differently. In this example, we scheduled nodes differently at different layers and we got a higher normalized sum-rate than that of MIR scheduling. We will formally define this scheduling in the following subsection and we refer to it as Maximum Independent Link (MIL) scheduling. A general lower-bound on the performance of MIR scheduling has also been derived in Appendix A.

*B. Maximum Independent Link (MIL) Scheduling*

MIL scheduling is defined as follows.

**Definition 6. MIL Scheduling:** *We assign a set, $\mathcal{J}(e)$, to each link $e$, containing all the indices of the source-destination pairs, that have that link on a route between them. We form a new set, called the color-index set, associated with each link that pairs up each source-destination index with a unique color from a set of $t$ colors. Therefore, such a set would look like $\{(color_i, j)\}$, where $i = 1, 2, \ldots, t$ and $j \in \mathcal{J}(e)$.*

*A coloring is valid, if the union of color-index sets of all incoming edges at a node, does not contain $(color_i, j)$ and $(color_i, m)$, where $j \neq m$. In other words, the transmitting node of a link does not interfere with any other link that has one of the same colors paired up with a different index, and the receiving node of an edge does not get interference from any other link that has one of the same colors paired up with a different index.*

*Each source-destination pair has a certain capacity when all the other source-destination pairs are silent. MIL scheduling is defined as scheduling the links using a valid coloring with $t$ colors, such that each source-destination pair achieves*

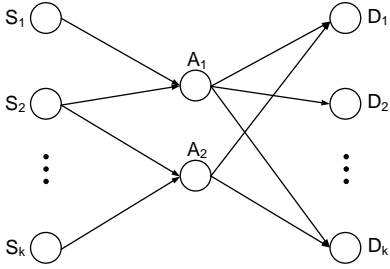

Fig. 7. A $k \times 2 \times k$ network.

its capacity over $t$ time slots. In fact, each color represents a time slot and in the time slot associated with $color_i$, all the links that have $color_i$ will be activated, and the source-destination pair that is paired up with that color on a link will use it for communication. The optimal MIL scheduling is the one that requires the minimum number of colors $t$.

We will show that MIL scheduling is optimal for a class of networks, namely $k \times 2 \times k$ networks, defined as follows.

**Definition 7.** *A $k \times 2 \times k$ network is a two-layer network with $k$ sources on one side, $2$ relays in the middle, and $k$ destinations on the other side. Note that there should exist at least one path from $S_i$ to $D_i$ in this network, $i = 1, 2, \ldots, k$, see Figure 7.*

**Theorem 6.** *The normalized sum-capacity of a $k \times 2 \times k$ network with $1$-local knowledge, is*

$$\alpha^* = \begin{cases} 1 & k = 1 \\ \frac{1}{\text{maximum degree of nodes in G}} & k \geq 2 \end{cases} \quad (5)$$

*where the maximum degree is defined as $\max\{d_i^{in}, d_i^{out}\}$, $i = 1, 2, \ldots, n$.*

*Proof:*
**Converse:** The result for $k = 1$ is trivial. Note that, in a $k \times 2 \times k$ network, the maximum degree happens at one of the relays for $k \geq 2$.

Suppose this maximum degree happens in the first layer, and without loss of generality assume it is the in-degree at relay $A_1$. Divide the sources into 3 disjoint subsets. Subset $V_i$ consists of all the sources that are only connected to relay $A_i$, $i = 1, 2$, and $V_{12}$ consists of the sources that are connected to both relays.

Our goal is to derive an upper bound on the normalized sum-rate of this network. Consider the corresponding destinations of the sources in $V_1$. Any such destination is either connected to relay $A_1$ or to both relays. If it is connected to both, then set the channel gain of any link from relay $A_2$ equal to zero. Follow the similar steps for members of $V_2$ and assign channel gain of $n$ to all other links in the network.

Relay $A_1$ should be able to decode all the messages coming from sources in $V_1$, since it has more information about messages than the intended destination of each message. A similar claim is valid for relay $A_2$, *i.e.* it should be able to decode all the messages coming from $V_2$. Relays $A_1$ and $A_2$ will decode messages coming from members of $V_1$ and $V_2$ respectively, and remove them from their received signals.

Now, the relays should be able to decode the rest of the messages together. However, since all the channel gains are the same, they have the same signal after removing the previously decoded messages. As a result each relay should be able to decode the remaining messages, which means relay $A_1$ is able to decode all the messages from $V_1$ and $V_{12}$ using its $n$ signal levels, note that $d_{A_1}^{in} = |V_1| + |V_{12}|$.

Since we assume 1-local knowledge at the sources, to achieve a normalized sum-rate of $\alpha$, each source should transmit at a rate greater than or equal to $\alpha n - \tau$. This together with the MAC capacity at relay $A_1$, gives us

$$d_{A_1}^{in}(\alpha n - \tau) \leq n \Rightarrow (d_{A_1}^{in}\alpha - 1)n \leq d_{A_1}^{in}\tau \quad (6)$$

since this has to hold for all values of $n$ where $\alpha$ and $\tau$ are independent of $n$, we get $\alpha \leq \frac{1}{d_{A_1}^{in}}$.

A similar argument is valid for the case where maximum degree happens in the second layer of the network, therefore we get

$$\alpha \leq \frac{1}{\text{maximum degree of nodes in G}} \quad (7)$$

this completes the proof of the converse.

**Achievability:** In a $k \times 2 \times k$ network, any link can at most be on one route. MIL scheduling works as follows. Assume that $d_{A_1}^{in}$ is the maximum degree in the graph.

Consider the first layer, we pick one link corresponding to a member of $V_1$ and one corresponding to a member of $V_2$ randomly, and we assign to both of them the same color. We keep picking two new links and assign them an unused color till there is no more link connected to a member of $V_2$. We then assign to each remaining link connected to a member of $V_1$ a new unique color. Any member of $V_{12}$, has either one link or two links associated with its index. In either case we assign a new unique color to the associated link(s). This way, we need $d_{A_1}^{in}$ number of colors.

Following the same steps and we will need $\max\{d_{A_1}^{out}, d_{A_2}^{out}\}$ colors to color all the links in the second layer. The total number of colors is therefore equal to the maximum degree of the graph, which alongside with MIL scheduling, completes the achievability proof. Note that the assumption that $d_{A_1}^{in}$ is the maximum degree in the graph, can be considered for any other degree and the same reasoning is valid there. Our achievability scheme holds for $\tau = 0$. ∎

The result presented as Theorem 6 will be extended to $k \times 2 \times \ldots \times 2 \times k$ networks, defined as follows.

**Definition 8.** *A $k \times 2 \times \ldots \times 2 \times k$ network is a layered network with $k$ sources in the first layer, $k$ destinations in the last layer, and $2$ relays in each layer in between. Any link in this network is either from a source to a relay, from one relay to another, or from a relay to a destination. Note that relays in the same layer are not connected to each other.*

**Theorem 7.** *The normalized sum-rate capacity of a $k \times 2 \times \ldots \times 2 \times k$ with $1$-local knowledge is*

$$\alpha^* = \begin{cases} \frac{1}{\text{maximum degree of nodes in } \mathsf{G}} & d_{A_i}^{out} = 1 \\ \frac{1}{k} & \text{otherwise} \end{cases} \quad (8)$$

*where as before the maximum degree is defined as* $\max\{d_i^{in}, d_i^{out}\}$, $i = 1, 2, \ldots, |\mathsf{V}|$.

Proof of this theorem is presented in Appendix B.

Again, it is quite interesting to see whether MIL scheduling is always optimal or not. The following example, will lead us to the answer. We will refer to the network depicted in Figure 8 as the two-layer $(3, 2)$ folded-chain network. We will see that the normalized sum-capacity of this network is $\frac{1}{2}$ and cannot be achieved by MIL scheduling. Note that MIL scheduling achieves only $\alpha = \frac{1}{3}$. This example is significant, in the sense that it depicts a case where network-coding is required. We will achieve the normalized sum-capacity by using repetition coding at the sources and linear network-coding at relays.

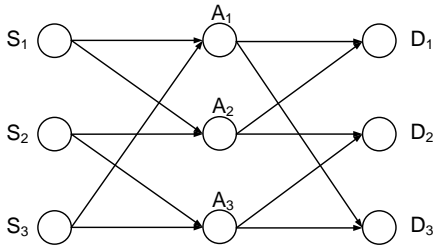

Fig. 8. *Two-layer $(3, 2)$ folded-chain network.*

See Theorem 9 for the fact that $\alpha \leq \frac{1}{2}$. We achieve $\alpha = \frac{1}{2}$ as follows.

First note that each source-destination pair has a diamond network, when other source-destination pairs are silent. In this diamond network, there exists an optimal achievability scheme, as shown in Figure 9 for pair 1. We will give each source-destination pair a chance to implement its optimal strategy for the diamond network, in the two-layer $(3, 2)$ folded-chain network over two time slots. In the first layer, we use repetition coding at the sources as in section III, which will provide all relays with the messages they require to create their input to the channel as if they are in any of the diamond networks, this happens in two time slots.

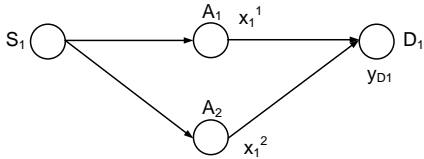

Fig. 9. *The diamond network of pair 1 when other source-destination pairs are silent.*

In the second layer, the achievability scheme works as follows. In each time slot, we will serve one destination completely. This way one relay is free at each time slot and the goal is to use this relay to serve the remaining the destination that will not be served completely in any time slot. See Table I for details of a possible scheme.

In time slot 1, Relays $A_1$ and $A_2$ serve Destination 1 and Relay $A_3$ sends its signal to Destination 3 alongside with its signal for Destination 2. Note that Destination 2, does not have to be on in the first time slot. In second time slot, Relays $A_2$ and $A_3$ serve Destination 2 and Relay $A_1$ combines its signal for Destination 3 with its previous sent signal and transmit it. At the end of second time slot, if Destination 2 combines all its received signals, the interference will cancel out and it will get its message, *i.e* $y_3^1 \oplus y_3^3$, interference-free. In this table, $y_i^{j'}$ represents the version of the message for Destination $i$ through Relay $A_j$, $x_i^j$, received at a destination different from $i$, and $y_i^j$ represents $x_i^j$ as received at Destination $i$.

TABLE I
SECOND LAYER ACHIEVABILITY SCHEME FOR THE NETWORK DEPICTED IN FIGURE 8.

| Destination id | time slot # 1 ($D_1$ being served) | time slot # 2 ($D_2$ being served) |
|---|---|---|
| $D_3$ | $y_1^{1'} \oplus y_2^{3'} \oplus y_3^3$ | $y_3^1 \oplus y_2^{3'} \oplus y_1^{1'}$ |
| | Relay $A_3$ serves $D_3$ | Relay $A_1$ serves $D_3$ |

As described in the last example, in some cases we need to combine MIL scheduling with linear network coding, in fact we can generalize the previous example as follows.

**Definition 9.** *A two-layer $(k, d)$ folded-chain network is a two-layer network with $k$ source-destination pairs and $k$ relays in the middle. Each source-destination pair $i$ has $d$ disjoint paths, through relays $i, i+1, \ldots, i+(d-1)$ **mod** $k$, $i = 1, 2, \ldots, k$. See Figure 10 for a depiction.*

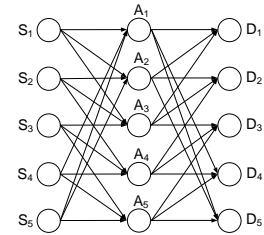

Fig. 10. *A two-layer $(5, 3)$ folded-chain network.*

**Theorem 8.** *The normalized sum-capacity of a two-layer $(k, d)$ folded-chain network with $1$-local view is $\alpha^* = \frac{1}{d}$, and is achieved by CMIL scheduling.*

Suppose in the first layer all channel gains are equal to $n$ and in the second layer, all channel gains are equal to zero, but the ones from $A_i$ to $D_i$, $i = 1, 2, \ldots, k$, which are equal to $n$. With this setting the idea developed to upper bound the normalized sum-rate in Theorem 3, is valid here. The achievability is a simple extension of the one developed for the example of two-layer $(3, 2)$ folded-chain network and is omitted in the paper.

## V. UPPER BOUND

In this section, we will develop general upper bounds on the normalized sum-capacity. One immediate upper bound on the normalized sum-capacity of an acyclic network with 1-local view is presented below.

**Theorem 9.** *In a k-user acyclic network with* 1-*local view, if there exists a path from* $S_i$ *to* $D_j$, $i \neq j$, *then the normalized sum-capacity is upper bounded by* $\frac{1}{2}$.

*Proof:* Consider a path, $P_{ij}$, from $S_i$ to $D_j$, $i \neq j$. Assign channel gain of $n$ to all edges in this path. For each source-destination pairs $i$ and $j$, create exactly one path from the source to destination with all channel gains equal to $n$. Assign channel gain of $0$ to all remaining edges in this network. See Figure 11 for a depiction.

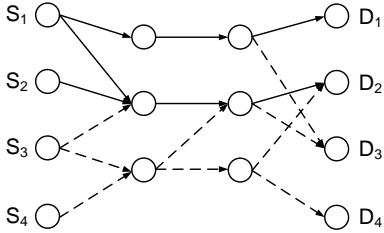

Fig. 11. *A path exists from* $S_1$ *to* $D_2$; *all solid edges have capacity n and the rest have capacity* 0.

In order to guarantee a normalized sum-rate $\alpha$, each source should transmit at a rate $\geq \alpha n - \tau$. Sppose this rate is feasible, therefore $D_j$ is able to decode $W_j$ and it will remove it from the received signal. The remaing signal is exactly the signal that $D_i$ receives, and since $D_i$ should be able to decode $W_i$, $D_j$ can decode it as well. Hence, the MAC capacity at $D_j$ gives us

$$2\alpha n - 2\tau \leq n \Rightarrow (2\alpha - 1)n \leq 2\tau \qquad (9)$$

since this has to hold for all values of $n$ where $\alpha$ and $\tau$ are independent of $n$, we get $\alpha \leq \frac{1}{2}$. ∎

In Section III, the binary model and the symmetric capacity were introduced. Based on those ideas, we will develop a general upper bound for acyclic network with 1-local view. Assume that all the channel gains in the network have equal capacity, $n$, however, we still assume the assumption of 1-local view at sources. In other words, each source is only aware that all the edges that are on a path form it to a destination, have equal capacity, $n$, but it has no information about the channel gains of other edges.

Suppose that in the network described above, the sum-capacity is equal to $\alpha^*_{sym}$. Then, since the channel gains in this network form a realization of network state, we have $\alpha^* \leq \alpha^*_{sym}$. The significance of this result is that, in all the networks we have considered so far, we have $\alpha^* = \alpha^*_{sym}$. However, unlike the single-layer case, where the achievability scheme of binary model could be generalized, in this case the achievability scheme can be quite different, as in the example depicted in Figure 8. An interesting direction is to figure out for what classes of networks this upper bound is tight.

## VI. FUTURE WORK AND CONCLUSION

In this paper, we developed a new model for partial netwrok knowledge in acyclic wireless networks. Different schemes were studied and their performances were examined in several cases, in terms of the normalized sum-capacity. Deep connections between network topology, normalized sum-capacity and the achievability strategies were noticed, and we saw several motivating results. One of the important future directions is to extend our results to Gaussian networks.

In this paper, we have only studied cases with 1-local view, a major direction is to figure out how normalized sum-capacity increases as nodes learn more and more about the network. Moreover, we have considered the case that the nodes know the network-connectivity globally, but the actual values of the channel gains are only known for a subset of routes. An appealing extension in this case, would be to understand the effects of local knowledge about network connectivity on the capacity and develop distributed strategies to optimally route information with partial knowledge about network connectivity. It is also desired to develop scalable network coding strategies that only rely on local network information. As mentioned before, the very idea of developing fundamental information-theorectic based foundations for networks with partial information at nodes is in its beginning stages and as a result, there are many salient directions to work on.

## APPENDIX A
## A LOWER-BOUND ON THE PERFORMANCE OF MIR SCHEDULING

In order to establish a lower-bound on the performance of MIR scheduling, we need the following definition.

**Definition 10.** *The* **pair-adjacency graph** *of a network is formed by representing each source-destination pair by a vertex in a graph. We put an undirected edge between two vertices* $i$ *and* $j$ *in this graph, if and only if there exists a path from* $S_i$ *to* $D_j$ *or from* $S_j$ *to* $D_i$. *Note that, pair-adjacency graph is an extension of conflict graph introduced in section III.*

The maximum number of time slots required by MIS scheduling in a network, is equal to the $m$-fold chromatic number of the pair-adjacency graph, which is defined in section III. We will next establishe a lower-bound on the performance of MIR scheduling for two-layer networks as defined below.

**Definition 11.** *A* **two-layer network** *is a network with* $k$ *sources on one side, a number of relays in the middle, and finally* $k$ *destinations on the other side. Any edge in this network is either from a source to a relay or from a relay to a destination. See Figure 5(a) as an example.*

**Theorem 10.** *Consider a two-layer network where* $d$ *is the maximum degree of nodes, i.e.* $\max\{d_i^{in}, d_i^{out}\}$, $i =$

$1, 2, \ldots, |\mathsf{V}|$, *in* $\mathsf{G}$ *with* 1-*local knowledge. The normalized sum-rate that is achieved by MIR scheduling, satisfies*

$$\alpha \geq \frac{1}{\min(k, 2d(d-1)+1)} \quad (10)$$

*Proof:* If two sources are connected to the same relay, their corresponding vertices in the pair-adjacency graph are adjacent, the same is true for any two destinations. The maximum degree in the pair-adjacency graph due to first layer is at most $\min(k-1, d(d-1))$ and this is also true for the second layer. Therefore, the maximum vertex degree in the pair-adjacency graph would be $\min(k-1, 2d(d-1))$. Using *greedy* coloring we know that every graph can be colored with one more color than the maximum vertex degree. Therefore, we are able to color the vertices in this graph using at most $\min(k, 2d(d-1)+1)$ colors, which gives us the desired lower bound. Note that $\alpha = \frac{1}{k}$ can be achieved by doing equal time-division among the $k$ users. ∎

## APPENDIX B
## PROOF OF THEOREM 7

If each relay in a layer is only connected to one relay in the previous layer and one in the next layer, then for the converse, we set all the channel gains among relays equal to $n$, and all the other channel gains as in the proof of Theorem 6. Supoose the only outgoing edge from relay $A_i$ is connected to relay $A_j$, then they have exactly the same information and we can consider them as a single relay. Achievability, is the same as in theorem 6. This can be seen in Figure 12, here we can merge relays $A_1$ and $A_3$ and the same thing for relays $A_2$ and $A_4$.

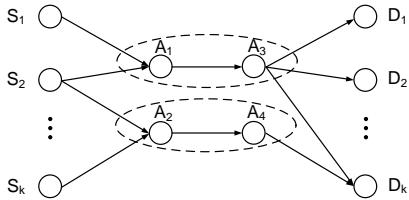

Fig. 12. A $k \times 2 \times \ldots \times 2 \times k$ network.

If the condition does not hold, we will have a relay that can decode all messages if we set the channel gains as in the previous case. From the MAC upper bound at such relay, we have $\alpha \leq \frac{1}{k}$. We can achieve this upper bound by TDMA and this completes the proof. Also note that the achievability scheme holds for $\tau = 0$.

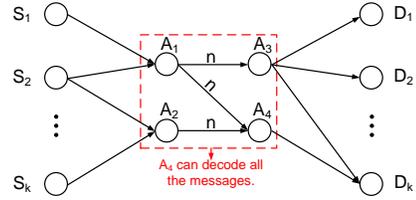

Fig. 13. A $k \times 2 \times \ldots \times 2 \times k$ network; $A_4$ can decode all the messages.